\documentclass[11pt]{article}

\usepackage{times}
\usepackage{url}
\usepackage{wrapfig}
\usepackage{psfrag}
\usepackage{url}
\usepackage{fullpage}
\usepackage{multirow}
\usepackage{amssymb}
\usepackage{amsmath}
\usepackage[dvips]{graphicx}
\usepackage[dvips]{color}

\usepackage{fancyvrb}
\usepackage[OT1,T1]{fontenc}

\CustomVerbatimEnvironment{myverbatim}{Verbatim}{
fontsize=\small,baselinestretch=0.9}

\usepackage{epsfig}

\newcommand{\Id}[1]{\ensuremath{\mathit{#1}}}

\newcommand{\realrange}[2]{\left[#1, #2\right]}

\newcommand{\unitrange}[2]{\realrange{0}{1}}

\newcommand{\Oh}[1]{\mathcal{O}\!\left( #1\right)}

\newcommand{\llabel}[1]{\label{\labelprefix:#1}}
\newcommand{\labelprefix}{} % later redefined using renewcommand

\newcommand{\labelcommand}{}
\newcommand{\captiontext}{}
\newsavebox{\buchalgorithmparam}
\newcounter{lineNumber}
\newenvironment{buchalgorithmpos}[3]{%
\renewcommand{\labelcommand}{#2}%
\renewcommand{\captiontext}{#3}%
\sbox{\buchalgorithmparam}{\parbox{\textwidth}{#3}}%
\begin{figure}[#1]\begin{center}\begin{code}\setcounter{lineNumber}{1}}{%
\end{code}\end{center}\caption{\llabel{\labelcommand}\captiontext}\end{figure}}

{\end{buchalgorithmpos}}

\newenvironment{code}{\noindent\it%
\begin{tabbing}%
\hspace{1.5em}\=\hspace{1.5em}\=\hspace{1.5em}\=\hspace{1.5em}\=\hspace{1.5em}\=%
\hspace{1.5em}\=\hspace{1.5em}\=\hspace{1.5em}\=\hspace{1.5em}\=\hspace{1.5em}\=%
\kill}{\end{tabbing}}

\newdimen\endofsize\endofsize=0.5em

\newcommand{\donotshow}[1]{}

\newcommand{\ignore}[1]{}

\newcommand{\qed}{\rule[-0.2ex]{0.3em}{1.4ex}}

\newcommand{\mbegin}{\{\ \ }

\newlength{\mleftindent}
\newlength{\mindent}
\newlength{\mboxwidth}

\newlength{\preprogramskip}
\newlength{\postprogramskip}
\newlength{\mexpwidth}
\newlength{\mexpindent}

\newlength{\proofpostskipamount}
               {\par\vspace{0.5ex}\noindent{\bf Proof #1:}\hspace{0.5em}}%
               {\nopagebreak%
                \strut\nopagebreak%
                \hspace{\fill}\qed\par\medskip\noindent}

\newcommand{\myurl}[1]{{\footnotesize \url{#1}}}

\def\usecs{{\mu}s}
\def\mysubsection{\vspace{-3mm}\paragraph*}

\newcommand{\kmreplace}[2]{{#2}}
\definecolor{dgreen}{rgb}{0,0.56,0} %darkgreen, xfig green4
\definecolor{dmagenta}{rgb}{0.56,0,0.56} %darkmagenta, xfig magenta4
\definecolor{dcyan}{rgb}{0,0.56,0.56} %darkcyan, xfig cyan4
\definecolor{dbrown}{rgb}{0.5,0.19,0} %darkbrown, xfig brown4
\definecolor{highlight}{rgb}{0,0,0}

\title{Engineering DFS-Based Graph Algorithms}
\author{Kurt Mehlhorn\thanks{MPI Informatik, Saarbr\"ucken, Germany} \and
        Stefan N\"aher\thanks{Universit\"at Trier, Germany, {\tt
naeher@uni-trier.de.}} \and
        Peter Sanders\thanks{Universit\"at Karlsruhe (TH), Germany, {\tt sanders@kit.edu}. 
                             Partially supported by DFG grant SA 933/3-1.}}
\date{}
\begin{document}

\maketitle\pagestyle{plain}

\begin{abstract} Depth-first search (DFS) is the basis for many efficient graph
algorithms. We introduce general techniques for \kmreplace{efficient
implementations}{the efficient implementation} of
DFS-based graph algorithms and exemplify them on three algorithms for computing strongly
connected components. The techniques lead to speed-ups by a factor of two to
three compared to the implementations provided by LEDA and BOOST. 
 We have obtained similar speed-ups for biconnected components
algorithms. We also compare the graph data types of LEDA and BOOST. 

\end{abstract}

%%%%%%%%%%%%%%%%%%%%%%%%%%%%%%%%%%%%%%%%%%%%%%%%%%%%%%%%%%%%%%%%%%%%%%
\section{Introduction}
\begin{figure}[ht]
\begin{center}
\epsfig{file=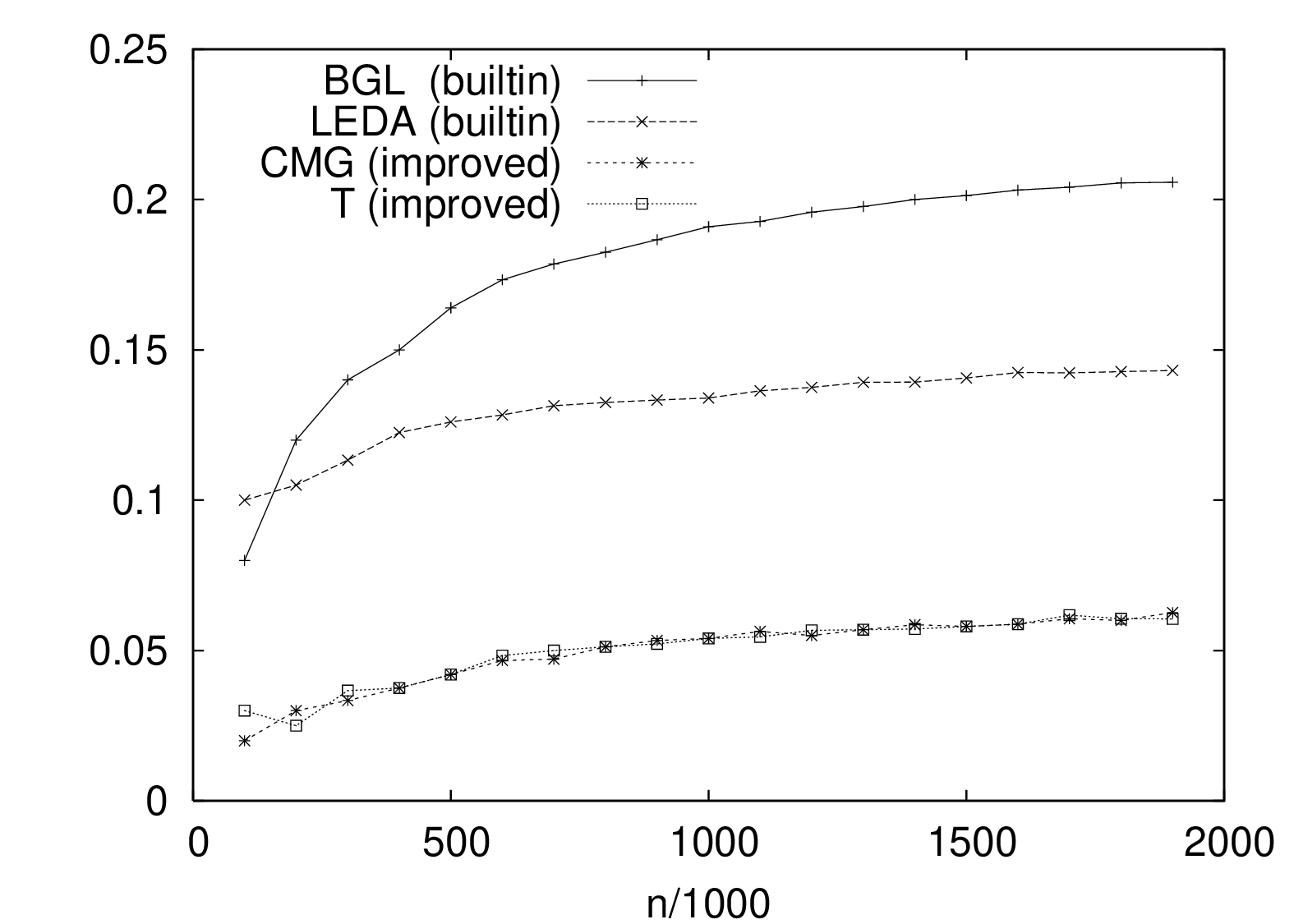,width=0.48\textwidth}\quad
\epsfig{file=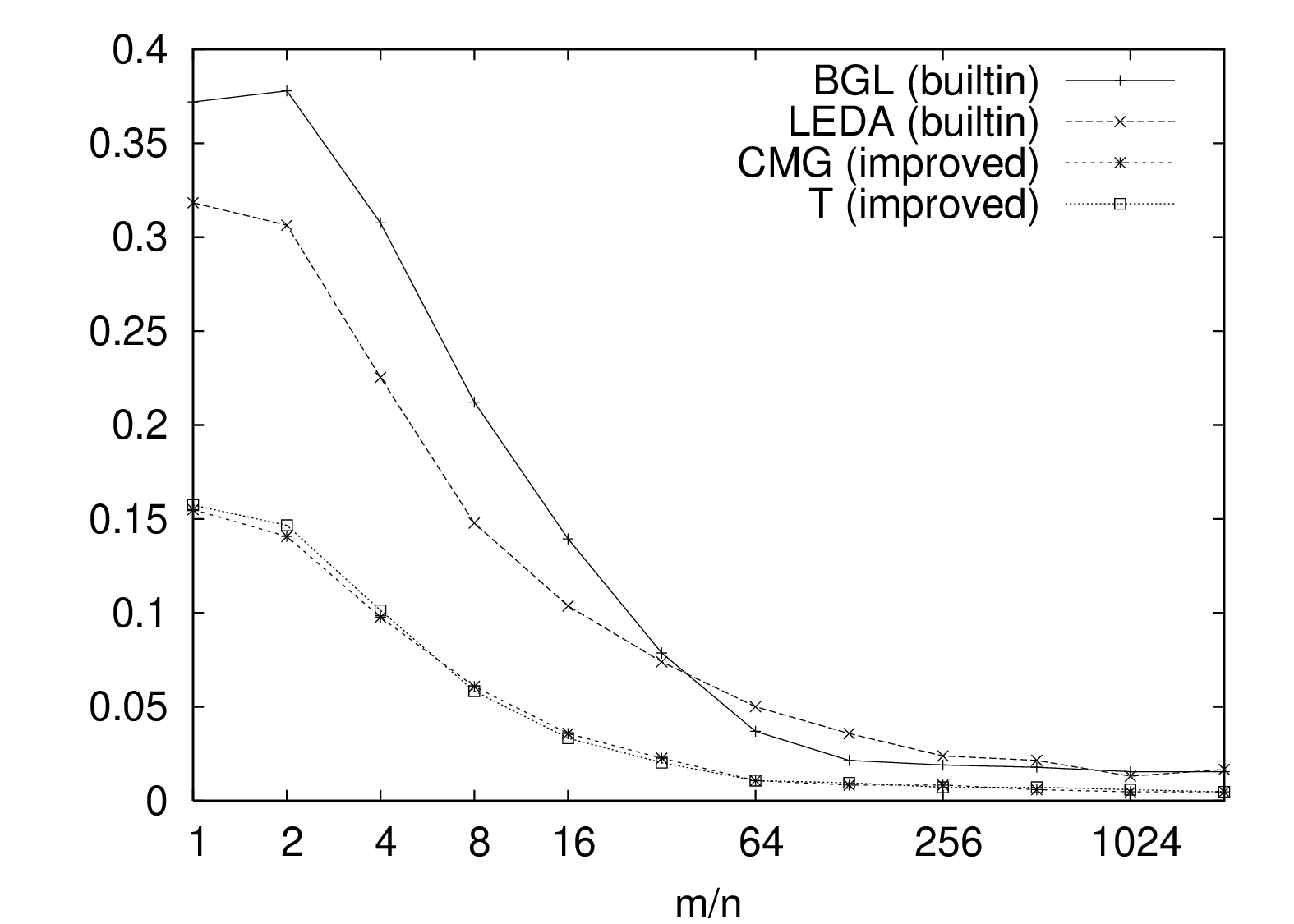,width=0.48\textwidth}
\end{center}
\caption{\label{fig:basic comparison} The running time per edge of four 
different SCC-implementations (in $\usecs$): the built-in implementations of 
BGL and LEDA and the optimized T- and CGM-programs. The figure on the left 
shows the times for different $n$ and fixed edge density $m/n = 10$ and the 
figure on the right shows the times for fixed $m = 2^{23}$ and different 
values of $m/n$. The running times of the optimized versions of T and CGM are 
essentially identical.}
\end{figure}

Depth-first search (DFS) is the basis for many efficient graph algorithms. We
introduce general techniques for the efficient implementation of DFS-based
graph algorithms and exemplify them on three algorithms for strongly connected
components of digraphs%
\footnote{This is the arxiv version of a paper that is actually from 2007 that we never came round to publish in a peer reviewed venue. However, we believe that the techniques warrant being published as a technical report.}. The techniques lead to speed-ups by a factor of two to three
compared to the implementations provided by LEDA~\cite{MehNae99,LEDA-AS} and BGL (BOOST Graph Library)~\cite{BOOST},
see Figure~\ref{fig:basic comparison}. The techniques 
apply to all DFS-based graph algorithms. We have already applied them to
biconnected components algorithms and obtained similar speed-ups. 

A \emph{strongly connected
component} (\emph{SCC}) of a directed graph (digraph) $G = (V,E) $ is a maximal subset of nodes $C$ with
the property that there is a directed path between any two nodes in
$C$. Computing SCCs means to label the nodes such that any two nodes belong to
the same SCC iff they have the same label. There are 
three different linear time algorithms for this
problem, all based on depth-first search: Tarjan's algorithm~\cite{Tar72}, the
Cheriyan-Mehlhorn-Gabow\footnote{This algorithm was first described by
Cheriyan-Mehlhorn~\cite{CM96} and later rediscovered by
Gabow~\cite{Gabow:scc}.} algorithm~\cite{CM96,Gabow:scc}, and the
Kosaraju-Sharir algorithm~\cite{Sha81}. We refer to these
algorithms as the T-, CMG-, and KS-algorithms, respectively. KS performs two
passes of DFS, one on $G$ and one on the reversed graph; the other two
algorithms rely on a single DFS with somewhat more complex bookkeeping. 

Modern processors are complex computing engines and the actual running time of
a program is mainly determined by three factors: (1) the number of instructions
executed, (2) efficient use of the processor pipeline, and (3) efficient use of
the cache memory. In particular, cache faults carry a high penalty. 

We \kmreplace{ will}{ shall} next sketch two of our speed-up techniques. 
Graph algorithms assemble information about the input graph during their
execution. Part of this information is associated with nodes, e.g., for a node
we may record whether DFS has reached it, 
the DFS-number, the number of the component
containing it, and so on. It is natural, to reserve separate storage
locations for the different pieces of information. However, this is frequently
wasteful and leads to an unneccesarily large memory footprint and hence an
unneccesarily large number of cache faults. The different node labels are
typically relevant at different times during execution and hence can be
combined into a single location. For example, node label zero may indicate that
DFS has not reached a node yet, a negative node label may indicate that the
node has been reached and record the (negated) DFS-number, and a positive node
label may indicate that the node is completed and has been assigned to an
SCC. Another example is that a particular label is only relevant while the
DFS-call for the node is active. Then the information is best stored on the
recursion stack. 

DFS only needs access to the edges leaving a node. Therefore, a very simple
graph representation suffices. All edges leaving a node a stored in a common
array. The arrays for different nodes may be combined into a single array. The
static graph types of BGL and LEDA use this representation. On this
representation,
\kmreplace{ scanning the graph may incur up to $2n
+ m/B$ cache faults, one for each call, one for each return, and one for each
block of $B$ edges. Here $B$ is the size of a cache line measured in number of
edges that fit into a cache line. 
We give an alternative implementation where the number of cache faults lies
between $n + m/B$ and $n +3m/B$.}
{ DFS performs $2n$ random accesses into this
array, one for each call (when the first edge out a newly reached node is accessed) and one
for each return (when control returns to a node and the scan of its out-edges is resumed). The
first kind of random access usually leads to a cache fault except in lucky
circumstances when the edge was brought into cache during the scan of another
node; the second kind of random access may or may not lead to a cache fault
depending on whether the next edge out of a node is still in cache when control
returns to the node. We give an alternative implementation that requires only
one random access to the edge array of node.} 
The edges emanating from a node are pushed onto a stack of edges
when the node is first encountered by DFS. Then the next edge to be explored is
always on the top of the stack and there is no cache fault when control returns
from a recursive call and resumes the scan of an adjacency array. 

Besides the technical contribution, our paper should also be useful teaching
material for a course in algorithms or algorithm engineering, as it shows how
careful analysis of even a simple program can lead to significant improvements
in running time. The forthcoming textbook~\cite{MehSan07} contains
implementation notes that hint towards efficient implementations. In fact, the
current paper grew out of the implementation notes for the chapter on graph traversal.

The structure of this paper is as follows. In Section~\ref{s:algorithms} we
review the algorithms and in Section~\ref{s:improvements} we discuss the
improvement techniques. Section~\ref{s:implementation} gives implementation
details and Section~\ref{s:experiments} describes and discusses the
experiments. Finally, Section~\ref{s:conclusion} offers a short conclusion. The
Appendix shows some sources.

\newcommand{\dfs}{{\mathit{dfs}}} 
\newcommand{\DFS}{{\mathit{DFS}}}

\section{The Basic Algorithms}\label{s:algorithms}

All SCC algorithms discussed in this paper are based on DFS.
They take a directed graph $G$ as input and compute a node array 
\Id{compNum} of integers indicating the component number of each vertex.
Figure~\ref{alg:dfs} gives a generic version of DFS from which
the different SCC algorithms can be instantiated by defining 
the operations \Id{Init}, \Id{TreeEdge}, 
\Id{NonTreeEdge}, \Id{FinishTreeEdge}, and \Id{FinishNode}.
Furthermore, all algorithms use a \emph{DFS-numbering} of the nodes
that gives the order in which nodes are visited. We \kmreplace{ will}{ shall} write
$u\prec v$ if $u$ has been visited before $v$. Sometimes they also need 
to know whether a node $v$ is \emph{finished}, i.e., whether 
the call $\dfs(v)$ has been completed.

\begin{figure}[ht]
\vspace{5mm}
\begin{center}
\begin{minipage}[t]{4.5cm}
\begin{tabbing}
{\bf procedure} \Id{DFS}($G$)\\
\ \ \ \=mark all nodes unvisited;\\
      \>\Id{Init};\\
      \>{\bf forall} $v \in V$ {\bf do}\\
      \>\hspace{5mm}\={\bf if} v is unvisited {\bf then}\\
      \>      \>\hspace{5mm} \Id{dfs}($v$);\\
      \>      \>{\bf fi};\\
      \>{\bf od};\\
\end{tabbing}
\end{minipage}\hspace{4cm}
\begin{minipage}[t]{4.5cm}
\begin{tabbing}
{\bf procedure} \Id{dfs}($v$)\\
\ \ \ \=mark $v$ visited;\\
  \>\Id{TreeEdge}(v);\\
  \>{\bf forall} $w\in V$ with $(v,w) \in E$ {\bf do}\\
  \>\hspace{5mm} \={\bf if} $w$ is unvisited {\bf then}\\
  \>      \>\hspace{5mm}\= \Id{dfs}($w$);\\
  \>      \>    \> \Id{FinishTreeEdge}($v,w$)\\
  \>      \>{\bf else}\\
  \>      \>    \> \Id{NonTreeEdge}($v,w$);\\
  \>      \>{\bf fi};\\
  \>{\bf od};\\
  \>\Id{FinishNode}($v$);
\end{tabbing}
\end{minipage}
\end{center}
\caption{Depth-first search of a directed graph $G = (V,E)$}
\label{alg:dfs}
\end{figure}

\ignore{All SCC algorithms discussed in this paper are based on DFS.
They take a directed graph $G$ as input and compute a node array 
\Id{compNum} of integers indicating the component number of each vertex.
Figure~\ref{alg:dfs} gives a generic version of DFS from which
the different SCC algorithms can be instantiated by defining 
the operations \Id{Init}, \Id{TreeEdge}, 
\Id{NonTreeEdge}, \Id{FinishTreeEdge}, and \Id{FinishNode}.
Furthermore, all algorithms use a \emph{DFS-numbering} of the nodes
that gives the order in which nodes are visited. We \kmreplace{ will}{ shall} write
$u\prec v$ if $u$ has been visited before $v$. Sometimes they also need 
to know whether a node $v$ is \emph{finished}, i.e., whether 
the call $\Id{dfs}(v)$ has been completed.}

\mysubsection{The algorithm of Kosaraju-Sharir:}
First, use $\Id{DFS}(G)$ to compute the order in which nodes are 
finished (\emph{finishing times}). Second, mark all nodes as unvisited,
traverse them in order of decreasing finishing time, and call $\Id{dfs}(v)$ 
on the reverse graph $G_{\mathrm{rev}}$ for each still unvisited node $v$. 
Each call will visit exactly a strongly connected component of $G$.

\begin{figure}[ht]
\vspace{5mm}
\begin{center}
\begin{minipage}[t]{4.5cm}
\begin{tabbing}
\Id{Init}:\\
\ \ \=\Id{sccCount} := 0;\\
    \>\Id{open} := empty stack;\\
\\
\Id{TreeEdge}($v$):\\
    \>\Id{lowPoint}[$v$] := $v$;\\
    \>\Id{open}.push($v$);\\
\\
\Id{FinishTreeEdge}($u,v$):\\
    \>{\bf if} \Id{lowPoint}[$v$] $\prec$ \Id{lowPoint}[$u$] {\bf then}\\
    \>\hspace{5mm}\=\Id{lowPoint}[$u$] := \Id{lowPoint}[$v$];\\
    \>{\bf fi};\\
\end{tabbing}
\end{minipage}\hspace{1.5cm}
\begin{minipage}[t]{4.5cm}
\begin{tabbing}
\Id{NonTreeEdge}($u,v$):\\
\ \ \={\bf if} $v$ open {\bf and} $v \prec$ \Id{lowPoint}[$u$]\ \ {\bf then}\\
    \>\hspace{5mm}\=\Id{lowPoint}[$u$] := $v$;\\
    \>{\bf fi};\\
\\
\Id{FinishNode}($v$):\\
    \>{\bf if} \Id{lowPoint}[$v$] = $v$ {\bf then}\\
    \>        \>{\bf repeat}\ \=$u$ := \Id{open}.pop();\\
    \>        \>              \>\Id{compNum}[$u$] := \Id{sccCount};\\
    \>        \>{\bf until}   \>$u$ = $v$;\\
    \>        \>\Id{sccCount}++;\\
    \>{\bf fi};\\
\end{tabbing}
\end{minipage}
\end{center}
\caption{Instantiation of basic DFS operations for Tarjan's algorithm.}
\label{alg:tarjan}
\end{figure}

\mysubsection{The algorithm of Tarjan:}
Consider the execution of $\Id{DFS}(G)$ and use $G_c = (V_c,E_c)$ to denote the 
currently explored subgraph, i.e., $V_c$ comprises the visited nodes 
and $E_c$ comprises the explored edges. 
We call an SCC of $G_c$ \emph{open} if it contains an unfinished node (= a
node $v$ for which $\Id{DFS}(v)$ has not finished yet)
and \emph{closed} otherwise. We call a node $v$ {\em open}
if $v$ belongs to an open component and {\em closed} if it belongs to
a closed component. 
For every closed node $v$ the number of its SCC is already known
and stored in \Id{compNum[v]}.

The algorithm maintains a stack \Id{open} of all open nodes and stores for
every open node $v$ its so-called lowpoint in a
node array \Id{lowPoint} such that $\Id{lowPoint}[v]$ is the node
with the smallest dfs-number among all nodes that are reachable from $v$
by a path of tree edges followed by one non-tree edge 
including node $v$ itself. It it not difficult to see that
%\kmfrage{I reformulated the sentence because root is not defined yet} 
\kmreplace{a currently
completed node $v$ (at the end of $\Id{dfs}(v)$) is a root of an SCC
if and only if $v = \Id{lowPoint}[v]$.}
{the status of a SCC changes from open
to closed when at the end of $\Id{dfs}(v)$, we have $v = \Id{lowPoint}[v]$}. 
Then $v$ and all nodes with a larger
dfs-number can be removed from the stack \Id{open}. They form a new SCC. 
The basic operations are shown in Figure~\ref{alg:tarjan}.

\begin{figure}[htb]
\vspace{5mm}
\begin{center}
\begin{minipage}[t]{4.0cm}
\begin{tabbing}
\Id{Init}:\\
\ \ \=\Id{sccCount} := 0;\\
    \>\Id{roots} := empty stack;\\
    \>\Id{open} := empty stack;\\
\\
\Id{TreeEdge}(v):\\
    \>\Id{roots}.push($v$);\\
    \>\Id{open}.push($v$);\\
\end{tabbing}
\end{minipage}\hspace{1cm}
\begin{minipage}[t]{4.0cm}
\begin{tabbing}
\Id{NonTreeEdge}($u,v$):\\
\ \ \={\bf if} $v$ is open {\bf then}\\
     \>\hspace{5mm}\={\bf while} $v \prec$ \Id{roots}.top() {\bf do}\\
     \>            \>\hspace{5mm} \Id{roots}.pop();\\
     \>            \>{\bf od};\\ 
  \>{\bf fi};\\
\end{tabbing}
\end{minipage}\hspace{1cm}
\begin{minipage}[t]{4.0cm}
\begin{tabbing}
\Id{FinishNode}($v$):\\
\ \ \={\bf if} $v$ = \Id{roots}.top() {\bf then}\\
    \>\hspace{5mm}\=\Id{roots}.pop()\\
    \>            \>{\bf repeat}\ \=$u$ := \Id{open}.pop();\\
    \>            \>              \>\Id{compNum}[$u$] := \Id{sccCount};\\
    \>            \>{\bf until}   \>$u$ = $v$;\\
\>\ \>\Id{sccCount}++;\\
\>{\bf fi};\\
\end{tabbing}
\end{minipage}
\end{center}

\caption{Instantiation of basic DFS operations for Cheriyan-Mehlhorn-Gabow.}
\label{alg:cheriyan-mehlhorn}
\end{figure}

\mysubsection{The algorithm of Cheriyan-Mehlhorn-Gabow:} 
\kmreplace{The algorithm is
similar to Tarjan's algorithm. 
In every component, the node with the smallest DFS-number in the component 
is called the \emph{representative} or {\em root} of the component. 
The idea of the AGM-algorithm is to  maintain the 
strongly connected components of $G_c$ during the execution of $\Id{DFS}(G)$.}
{The idea of the AGM-algorithm is to  maintain the strongly connected 
components of $G_c$ during the execution of $\Id{DFS}(G)$. In every component, 
the node with the smallest DFS-number in the component is called the 
\emph{representative} or {\em root} of the component.}

%\kmfrage{I reformulated the next paragraph}

Open components are represented by two stacks \Id{open} and \Id{roots}.
The stack
\Id{open} contains all open nodes in order of increasing DFS number and 
the stack \Id{roots} conains all roots in order of increasing DFS number. In
this way, \Id{roots} is a subsequence of \Id{open} and partitions \Id{open}
into the open SCCs of $G_c$. 
With these definitions in place, the basic operations become very simple.
They are shown in Figure~\ref{alg:cheriyan-mehlhorn}.

Note that the total number of executions of the loops in these operations is
$\Oh{n}$ since each iteration pops data from a stack that experiences
at most $n$ push operations.
Moreover, the operations on the stacks will be very fast in practice,
since stacks have good cache locality.

\section{Performance Improvements}\label{s:improvements}

We \kmreplace{ will}{ shall} first describe our improvements for the CMG
algorithm, mostly because this is the algorithm for which we implemented them
first. Then we discuss modifications needed for the other algorithms. 
Our main concern will be large graphs where cache faults due
to random memory accesses are the main cost factor.

\begin{figure}[t]
\begin{center}
\epsfig{file=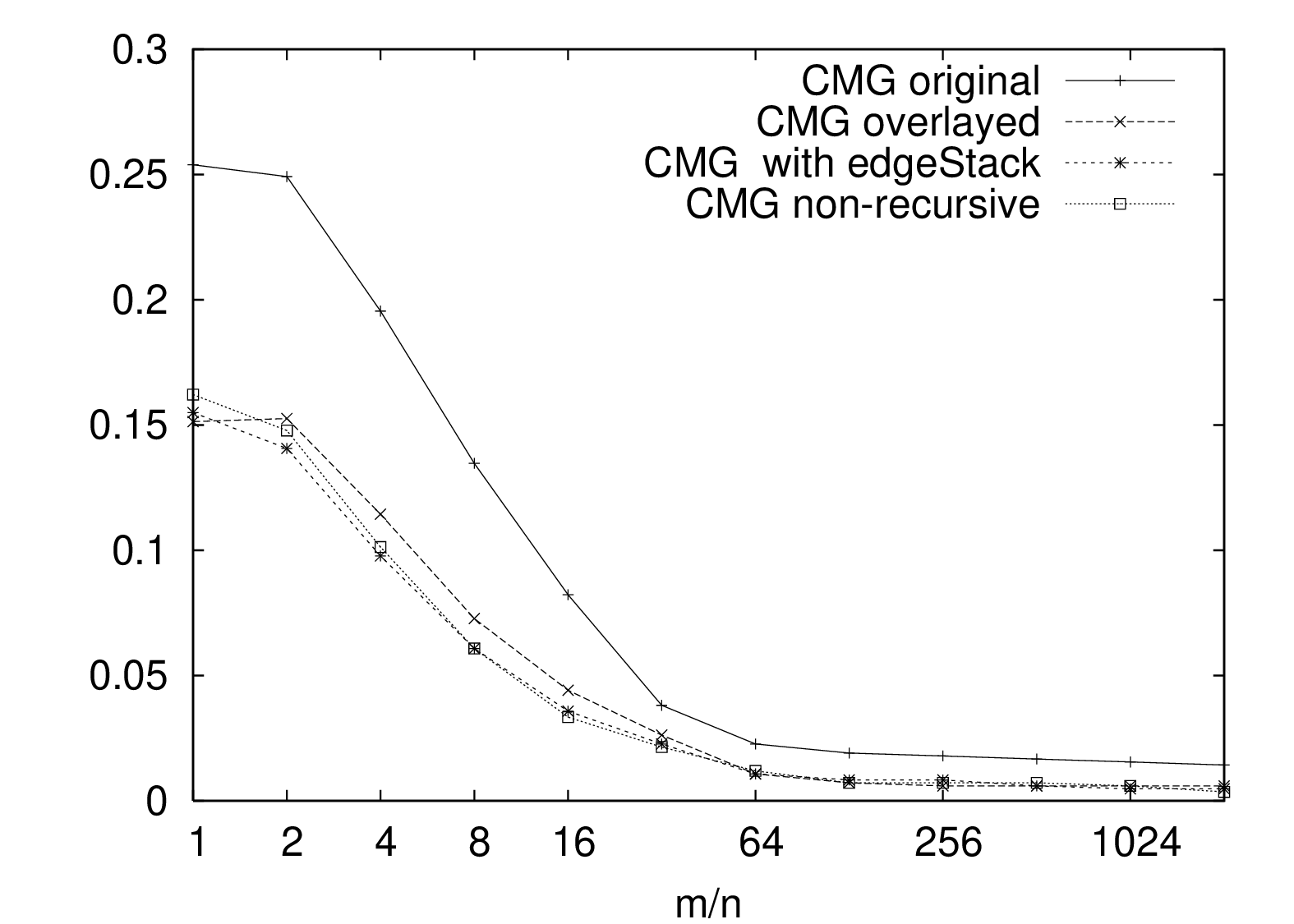,width=0.48\textwidth}\quad
\epsfig{file=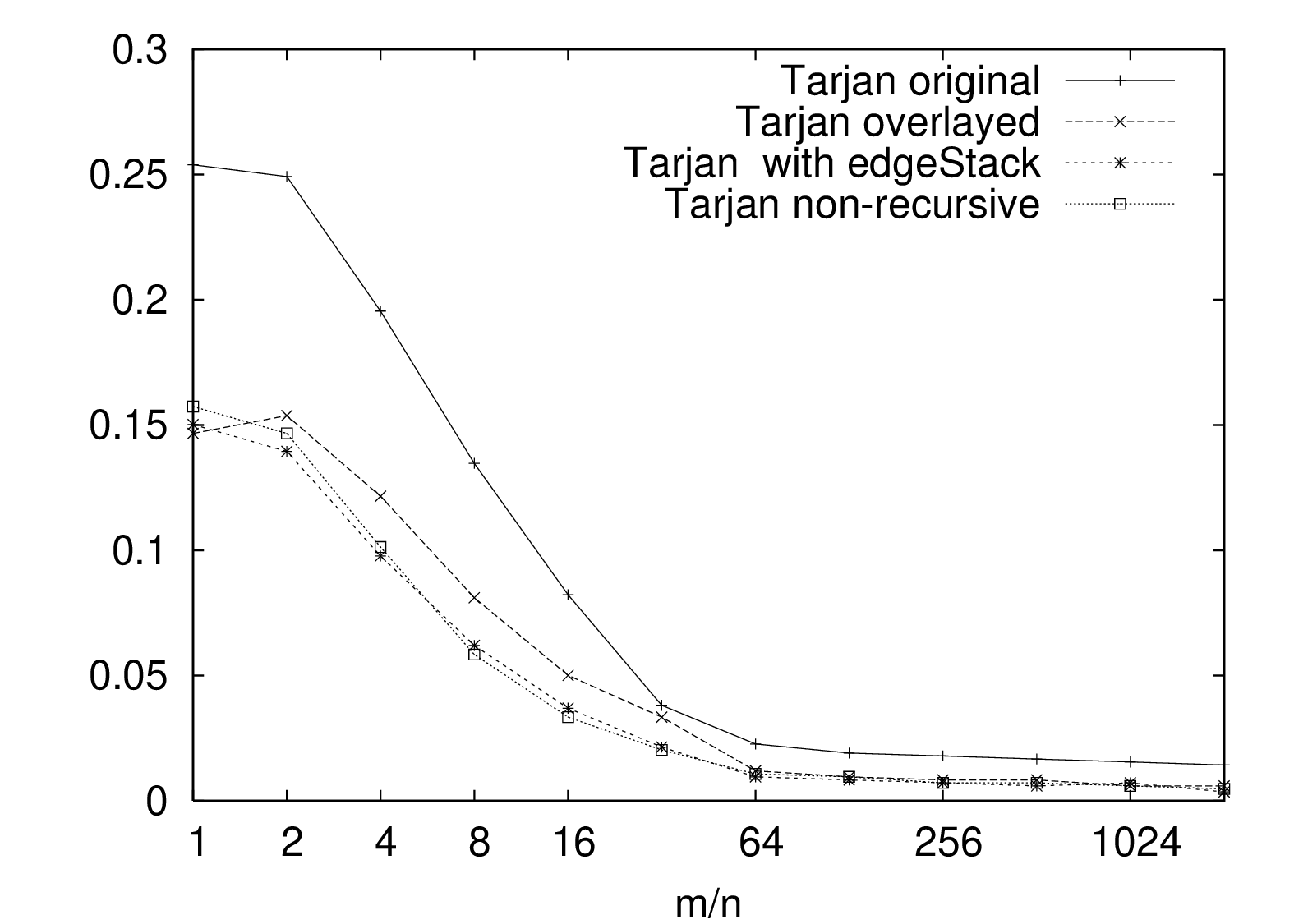,width=0.48\textwidth}
\end{center}
\caption{\label{fig:improvements} The performance improvements obtained for
CMG and T by the optimizations described in Section~\ref{s:improvements}. 
The time per edge is shown in $\usecs$ ; $m/n$ varies and $m = 2^{23}$. 
We turned on one optimization after the other.}
\end{figure}

\mysubsection{Overlayed Node Data:}
Besides the graph itself, CMG needs to store and access
four kinds of information for each node $v$: an indication whether $v$
is marked, an indication whether $v$ is open, something like a
DFS-number in order to implement `$\prec$', and, for closed nodes, the
representative of its component (the output). It seems that previous
implementations kept this information separate leading to significant
space overhead and many cache faults.  The array $\Id{compNum}$
suffices to keep this information. For example, if \Id{NodeId}s are
integers in $1..n$, $\Id{compNum}[v]=0$ indicates an unmarked
node. Negative numbers indicate negated DFS-numbers so that $u\prec v$
iff $\Id{compNum}[u]>\Id{compNum[v]}$.  This works because `$\prec$'
is never applied to closed nodes.  Finally, the test $w\in
\Id{open}$ simply becomes $\Id{compNum}[v]<0$. Gabow's description~\cite{Gabow:scc} of the
algorithms already describes some of these optimizations. 

\mysubsection{Graph Representation:}
Since the graph is static and only outgoing edges need to be known,
a very simple \emph{adjancency array} representation suffices.
Nodes are numbered $1$ through $n$.
A single edge array $E[1..m]$ stores the target nodes of all edges grouped
by source node. A vertex array $V[1..n+1]$ stores the starting position of 
the edges for each node so that $E[V[v]..V[v+1]-1]$ stores the
target nodes of the edges leaving node $v$ ($V[n+1]=m$).
This leaves an important choice for storing the node array \Id{compNum}:
Store it as a separate array or make \kmreplace{he}{the} vertices a record with two fields.
For sufficiently large $n$, the latter choice has better cache locality since
the adjancency information and the \Id{compNum} of a node are 
frequently accessed together. 

\mysubsection{Make the Common Case Fast:}
Except for very sparse graphs, the most frequently executed 
operation is \Id{NonTreeEdge}. This operation needs
to know the DFS-number of $\Id{roots}.\Id{top}()$. 
Hence, we redefine \Id{roots} to store DFS numbers rather than
NodeIDs of representatives. 
The only other access to the content of \Id{roots} (in \Id{FinishNode},
``$v=\Id{roots}.\Id{top}()$?'')
does not become much more expensive either, since
the DFS-number of $v$ can be stored on the recursion stack (i.e., in a local variable).

\mysubsection{Reducing Accesses to the Graph:}
DFS performs a sequential scan of the edges leaving a node $v$.  However,
interruptions by recursive calls can lead to additional cache faults
when accessing the adjancency list of $v$.  It therefore makes sense to
copy the adjacency list to the recursion stack once and for all when
it is first accessed. All remaining accesses can then be served from
the stack.  This is more cache efficient since stacks have very good
cache locality independent of the access pattern. A remaining problem
is that C++ does not support variable size arrays on the stack.
Hence, we use a separate stack for storing the copied adjacency arrays.
In a sense, we create a new representation of the graph with
a layout tuned for the order in which the nodes are accessed by DFS.

\mysubsection{Nonrecursive Implementation:}
Traditional wisdom tells us that recursive algorithms can be tuned
by making them nonrecursive. This is always possible by maintaining
a stack explicitly. This can be more efficient because
we have more control over content and representation of the stack.
However, so far our experiments show little difference. 
Apparently, modern compilers are good at keeping only things on
the stack that are really changing between the recursive calls.

\mysubsection{Tarjan's Algorithm:}

All optimizations above are also applicable to Tarjan's algorithm.
An additional optimization which was already 
discussed by Sedgewick in \cite{Sedgewick92} is as follows:
Low points do not have to be stored in a node array since lowpoint data
has to be available only for the current node $v$ and the neighbor node
$w$ that was just visited by a recursvive call $\Id{dfs}(w)$ (in \Id{FinishTreeEdge}).
It is sufficient to keep the current lowpoint of $v$ in a local variable
and to make $\dfs(w)$ return the lowpoint of $w$. 
This improvement makes Tarjan's algorithm more elegant. In fact, after the
optimization, the T- and the CMG-algorithm are quite similar. 
The latter has an additional stack where the former needs
an additional entry on the recursion stack.

\mysubsection{Kosaraju-Sharir:} Most of the above optimizations are also
relevant for this algorithm. The first pass can be simplified
to the extent that not even finishing times need to be stored in a
node array.  Instead, a \Id{FinishNode} operation simply pushes the NodeId
on a stack constituting the output. Computing the reverse graph is in
itself not so easy. It is comparable to sorting the edges by their
target node. Although cache efficient integer sorting algorithms might
help here, the cost is not negligible. Even if the input graph already
has reverse edges available, we still end up with two passes rather
than one pass. Within the passes, only the cheapest operations (stack
accesses,\ldots) can be saved compared to the one-pass
algorithms. The bottom line is that Kosaraju-Sharir does
not look promising for a high performance implementation. Its main
asset is its simplicity.

%%%%%%%%%%%%%%%%%%%%%%%%%%%%%%%%%%%%%%%%%%%%%%%%%%%%%%%%%%%%%%%%%%%%%%
\section{Implementation}\label{s:implementation}

For the implementation we used four different kinds of graph data 
structures: LEDA dynamic and static graphs, BOOST static graphs, and
a simple hand coded graph data structure.

In order to use the same code base for LEDA and BGL graphs we used
an adapter class that makes the typical LEDA iteration macros 
and node\_array syntax usable for BGL graphs as well. When working
with  LEDA graphs it is possible to choose whether the compNum array 
is stored externally as a {\em node\_array} or whether this data should be
stored directly in the node objects of the graph by using a {\em node\_slot}. 
The latter gives much better performance, in particular, when used together 
with a static graph structure. A detailed description of static graphs and 
node slots can be found in~\cite{NahZlo02}.

For the recursive versions of the code, it is necessary to work with a
unlimited system stacksize. On most operating systems the program
stacksize is limited by default to a very small quantity. This will
most definitely cause stack overflow errors for large inputs. The stacksize can
be set by using the limit command on Linux
({\tt limit stacksize unlimited}).

%%%%%%%%%%%%%%%%%%%%%%%%%%%%%%%%%%%%%%%%%%%%%%%%%%%%%%%%%%%%%%%%%%%%%%
\section{Experiments}\label{s:experiments}

For our experiments we use random graphs with varying $n$ and $m$. We performed
two sets of experiments, one for varying $n$ and fixed edge density $m/n = 10$
and one for fixed number of edges $m = 2^{23}$ and varying $n$. For all
algorithms under consideration, the number of instructions executed is almost independent of the
graph. Also, random graphs are not likely to be easy cases since random
edges imply many random memory accesses.  In the following discussion we
concentrate on large graphs with $2^{23}$ edges in order to see caching
effects clearly.  The implementation was done using gcc version 4.1 
using LEDA 5.0 and the Boost graph library version 1.34.1.  Optimization
level was -O2 producing 32 bit code. We report running times on one core of a PC with 2 GByte of memory and a 2.4 GHz Intel processor model Core2-Duo 6600 .
We obtained very similar results for AMD Opteron and SUN Sparc processors.

Figure~\ref{fig:basic comparison} compares the performance of our
best implementations for T and CMG with previous
implementations. 
We use the static graph data structure from LEDA and all the optimizations
outlined in Section~\ref{s:improvements}. The source codes can be found in the
appendix. We see that the tuned versions of T and CMG have essentially the same
performance and are significantly faster than the built-in implementations of
BGL and LEDA. We also see that the running time per edge decreases with
increasing edge density $m/n$. This is easily understood. The running time of
the algorithms depends linearly on $n$ and linearly on $m$, however the
dependence on $n$ comes with a larger factor of proportionality. Note, that
tree edges cause recursive calls (but there are only $n - 1$ of them) and that
only $n - 1$ non-tree edges can cause the merging of components in CMG. In T we
may have $\Theta(m)$ changes of lowpoint values, but these are assignments to
local variables and hence almost free. Since the dependency on $n$ comes with a
larger factor of proportionality, the running time per edge decreases as a
function of $m/n$ for fixed $m$. 

%\kmfrage{warum sehe ich auch fuer grosse m/n immer noch einen deutlichen 
%  Unterschied in Figure 1?}

Of course, it is difficult to explain the sources of differences when
the underlying algorithm, implementation details, and graph
representation are changed at the same time. So let us look at one
aspect after the other. Figure~\ref{fig:improvements} shows the effect of our
different optimizations applied to CMG and Tarjan's algorithm. 
In all cases, we use LEDA's static graphs to represent graphs. 
After showing the performance of the original algorithm, we
turn on one optimization after the other: First, the overlayed node data
improvement, then the edge stack to reduce accesses to the graph, and
finally a non-recursive version with both improvements.
The diagrams show that the first improvement reduces the running time
of both algorithms by a factor of about two, the edge stack gives another
improvement of about 10 percent, whereas the 
non-recursive implementation only yields a marginal improvement.

\begin{figure}[th]
\includegraphics[width=0.47\textwidth]{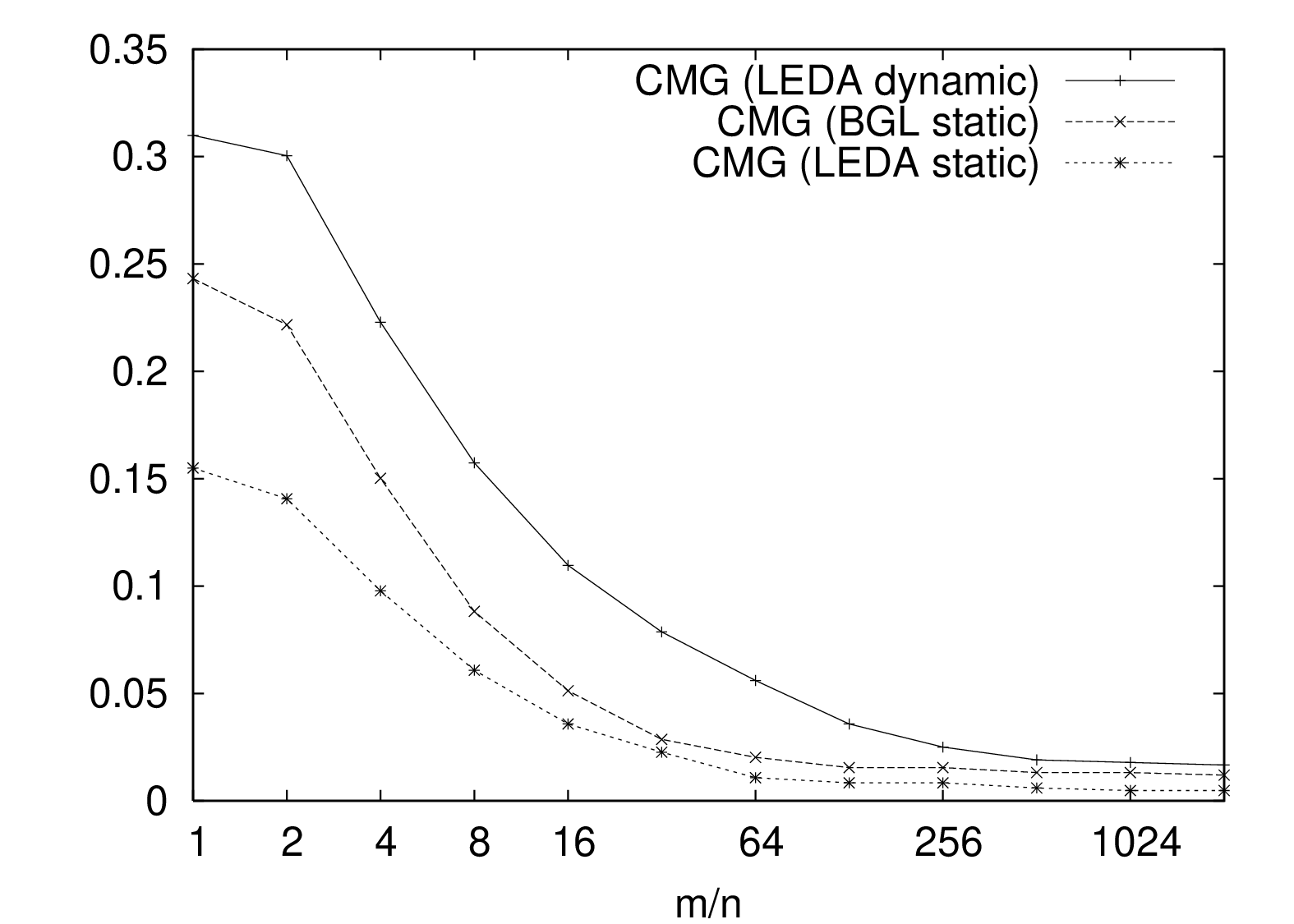}\quad\includegraphics[width=0.47\textwidth]{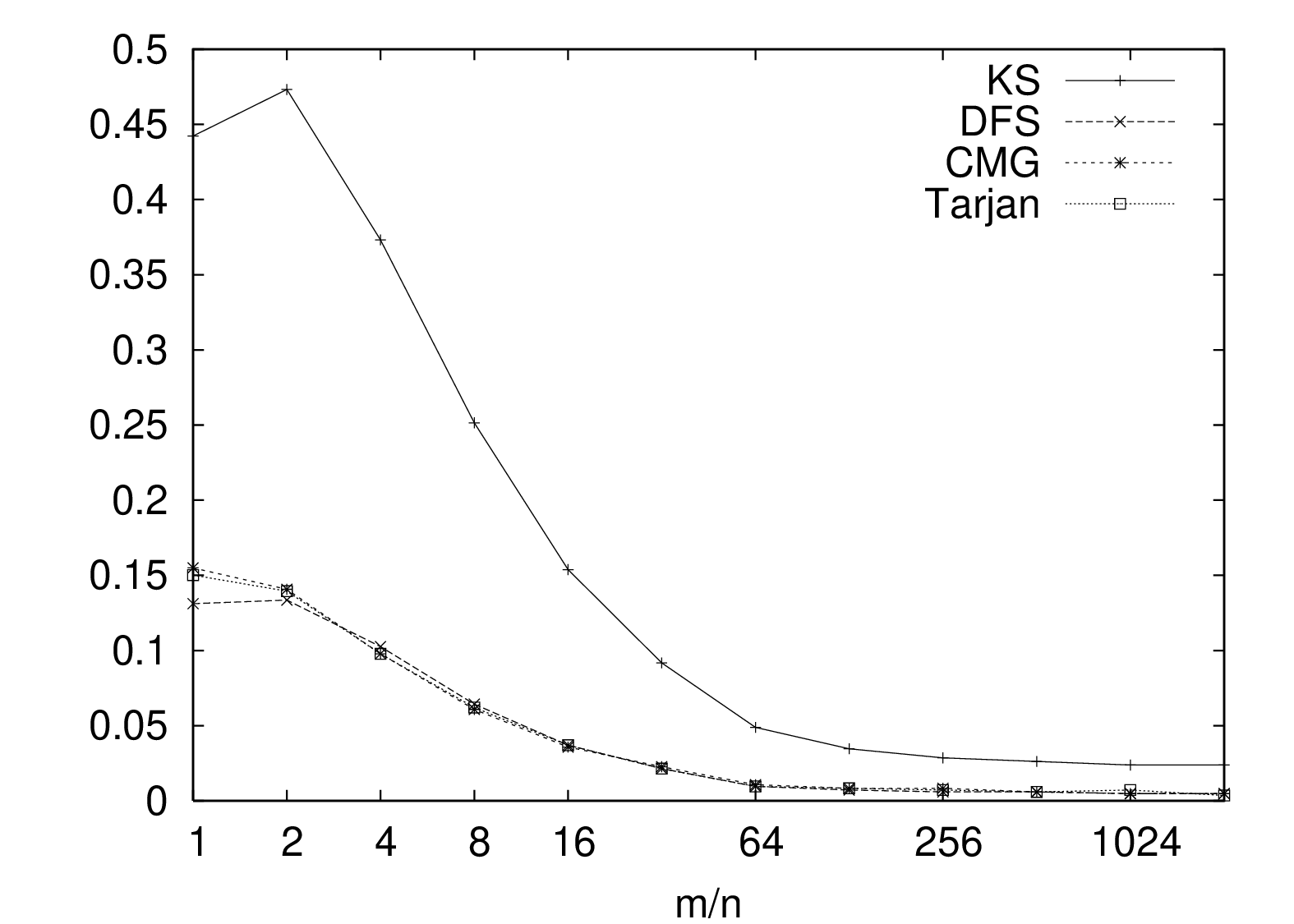}
\caption{The figure on the left shows running times of the tuned version of CMG-algorithm with BGL static and LEDA static and
dynamic graph representation. The figure on the right compares four algorithms:
Kosajaru-Sharir, Cheriyan-Mehlhorn-Gabow,
Tarjan, and simple DFS labelling}\label{fig:representation}\label{fig:algorithms}
\end{figure}

Figure~\ref{fig:representation} shows the effect of different graph representations on the running time
of tuned CMG. We see that LEDA's static graph data structure performs best, the
BGL implementation comes in second, and LEDA's dynamic graphs come in
last. This is to be expected as LEDA's dynamic graphs offer more constant time
update operations than BGL's graphs and LEDA's static graphs offer only a small
number of update operations. We explain the difference between LEDA's static
and dynamic graphs.

The static graph representation stores all edges in a
single array, the dynamic representation uses linked lists for the adjacency
lists.  The random graph generator ensures that the adjacency list of each
vertex is stored in consecutive locations. So the difference in running time is
due to the fact the dynamic representation uses substantially more space.

Finally, Figure~\ref{fig:algorithms} shows the running time of the tuned 
versions of all three algorithms together with the running time 
of a simple DFS-scan. We see that the DFS-scan takes almost as
much time as the two one-pass algorithms. A DFS-scan provides a 
lower bound for the execution time of any DFS-based algorithm. Thus the tuned
versions of T and CGM are essentially optimal and KS must take at least twice
the time of the other algorithms. In fact, it is worse since the reversed graph
must be computed in addition. 

\section{Conclusion}\label{s:conclusion}

\begin{figure}[ht]
\begin{center}
\epsfig{file=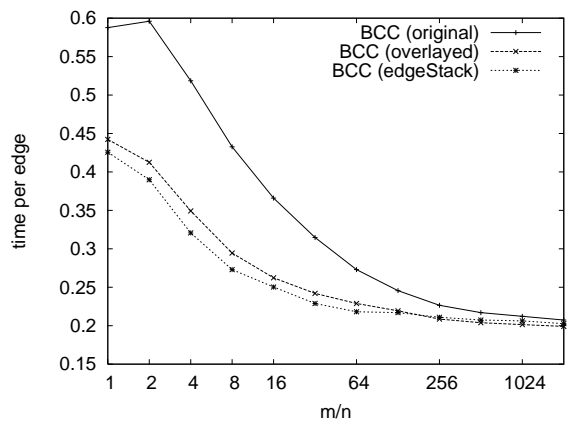,width=0.48\textwidth}
\end{center}

\caption{\label{fig:bicomps} The running time per edge of three 
different biconnected components implementations (in $\usecs$): 
the original algorithms (\cite{Tar72}) and the effect by applying 
the overlayed node data and edgeStack improvements one after the other.
The figure on the left shows the times for different $n$ and fixed edge density $m/n = 10$ and the figure on the right shows the times for fixed $m = 2^{23}$ 
and different values of $m/n$.}
\end{figure}

We have shown that implementation details have an (for us
unexpectedly) high impact on the performance of SCC algorithms. 
It is very likely that this extends to other algorithms based on
DFS. In particular, reducing accesses to the graph will work for any
DFS-based algorithm like DFS numbering, topological sorting, st-numbering, and others.
Overlaying the node data is more interesting for non-trivial
algorithms like biconnected components (\cite{Tar72}), 
triconnected components(\cite{HopTar73}),
or planarity testing (\cite{HopTar74}). 
Figure~\ref{fig:bicomps} shows that our techniques
also lead to substantial speed-ups for biconnected component algorithms. 

We have not yet looked on the influence of the graph structure on
performance. Sibeyn et al. \cite{SAC02} have looked at ways to 
handle large graph\kmreplace{}{s} using semiexternal algorithms where 
node descriptions fit into fast memory but edges \kmreplace{}{fit} only on slow external
memory with blocked access. Although the methods used there are too
complex to look promising for the cache-main-memory hierarchy,
some ideas like reordering nodes or adjacency lists might turn
to be relevant also for internal memory algorithms. 

We have tried prefetching instructions in order to hide memory access
latencies --- so far without success.

\newpage

\bibliographystyle{alpha}
\bibliography{diss,ref}

\begin{appendix}
\clearpage
\section*{Appendix: Selected Sources}
\subsection*{Tuned Cheriyan-Mehlhorn-Gabow}

\begin{myverbatim}
template <class graph_t> 
class STRONG_COMPONENTS_CMG {

int dfsCount;
int sccCount;

template<class compNumArray>
void dfs(const graph_t& G, node v, compNumArray& compNum, b_stack<node>& edgeStack,
                                                          b_stack<node>& open,
                                                          b_stack<int>&  roots)
{ int dfsNum = --dfsCount;
  compNum[v] = dfsNum;
  roots.push(dfsNum);
  open.push(v);

  int sz = edgeStack.size();

  edge e;
  forall_rev_out_edges(e,v) edgeStack.push(G.target(e));
  
  while (edgeStack.size() > sz)
  { node w = edgeStack.pop();
    int  d = compNum[w];
    if (d >= 0) continue;
    if (d == -1) 
      dfs(G,w,compNum,edgeStack,open,roots);
    else 
      while (roots.top() < d) roots.pop();
   }

  if (roots.top() == dfsNum) 
  { node u;
    do { u = open.pop();
         compNum[u] = sccCount;
        } while (v != u);
    roots.pop(); 
    sccCount++; 
   }
}

public:

template<class compNumArray> 
int run(const graph_t& G, compNumArray& compNum)
{ int n = G.number_of_nodes();
  int m = G.number_of_edges();
  b_stack<node> edgeStack(m);
  b_stack<int>  roots(n);
  b_stack<node> open(n);
  dfsCount = -1; sccCount = 0;

  node v;
  forall_nodes(v,G) compNum[v] = -1; 

  forall_nodes(v,G) 
   if (compNum[v] == -1) dfs(G,v,compNum,edgeStack,open,roots);

  return sccCount;
}
};
\end{myverbatim}

\clearpage
\subsection*{Tuned Tarjan}

\begin{myverbatim}
template <class graph_t> 
class  STRONG_COMPONENTS_TARJAN {

int dfsCount;
int sccCount;

template <class compNumArray>
int dfs(const graph_t& G, node v, compNumArray& compnum, b_stack<node>& edgeStack,
                                                         b_stack<node>& open)
{ int dfsNum = dfsCount++;
  int lowPoint = dfsNum;
  compNum[v] = dfsNum; 
  open.push(v);

  int sz = edgeStack.size();

  edge e;
  forall_rev_out_edges(e,v) edgeStack.push(G.target(e));

  while (edgeStack.size() > sz)
  { node w = edgeStack.pop();
    int d = compNum[w];
    if (d == -1) d = dfs(G,w,compNum,edgeStack,open);
    if (d < lowPoint) lowPoint = d;
   }

  if (dfsNum == lowPoint)
  { node u; 
    do { u = open.pop();
         compNum[u] = sccCount;
       } while (u != v);
    sccCount++;
   }

  return lowPoint;
}

public:

template <class compNumArray>
int run(const graph_t& G, compNumArray& compNum)
{
  int n = G.number_of_nodes();
  int m = G.number_of_edges();

  b_stack<node> edgeStack(m);
  b_stack<node> open(n);

  dfsCount = -(n+1); sccCount = 0;

  node v;
  forall_nodes(v,G) compNum[v] = -1;

  forall_nodes(v,G)   
     if (compNum[v] == -1) dfs(G,v,compNum,edgeStack,open);

  return sccCount;
}

};
\end{myverbatim}

\end{appendix}

\end{document}